\documentstyle[aps,pre]{revtex}
\begin{document} 
\draft 
\title{Reply to the Comment on ''Evidence for Neutrinoless Double
Beta Decay'' [Mod. Phys. Lett. {\bf A 16}(2001)2409]} 
\author{H.L.~Harney} 
\address{Max-Planck-Institut f\"ur Kernphysik, D-69029 Heidelberg,
         Germany\\} 
\date{\today} 
\maketitle 
\bigskip
\bigskip

The critics \cite{comment} of ref. \cite{KDHK} --- henceforth called
KDHK --- have listed nine objections. As coauthor of KDHK, I have 
advised the experimentalists in questions of statistics and want to 
discuss the nine points in the sequel. Part of the criticism is 
justified as one shall see.

\begin{enumerate}
\item ``There is no discussion of how a variation of the size of the ...
analysis window would affect the significance ...''

This is not true. The impact of the  analysis window was
qualitatively shown by KDHK via the  comparison of parts (a) and (b) 
of the Figs. 4--6.

Still I consider the size of the analysis window the most serious part
of the criticism. It is intimately linked to the following objection.

\item ``There is no relative peak strength analysis of all the 
$^{214}Bi$ peaks ...''

The analysis has been done under the assumption that the peaks showing up 
in the parts (a) of Figs. 4--6 of KDHK can be identified --- at least 
the ones closest to $Q_{\beta\beta}\, .$ If this is not so, the
significance decreases as one sees by comparing parts (a) and (b)
of the figures. By ``identification'', the agreement of position
and intensity of an observed structure with known $\beta$- or
$\gamma$-rays is meant.

According to present knowledge, the situation is not
as bad as the critics conclude from their Table 1 --- for two reasons:
(i) The expected rate of the weak $^{214}Bi$-lines is larger than the 
numbers in their table by a factorof about nine. This is due
to a problem in the normalization of Fig. 1 of \cite{HDM}; it is not 
the fault of the critics, cf. \cite{HVK}. (ii) Cascades of 
$\gamma$-rays emitted from 
a source that is close to the detector are partially summed by the
detector. As a consequence the intensities are not proportional to
the branching ratios given in \cite{FS}. A Monte Carlo simulation of 
the setup is needed to find the intensisies. 

The identification of the peaks would indeed have been the task of KDHK.
This means that the analysis window should have been taken wider.
Then the  analysis must include the simulation of the contaminating
peaks.

If the peaks at  energies other than $Q_{\beta\beta}$ cannot be identified 
by way of the simulation, the confidence on the possible structure
at $Q_{\beta\beta}$ will be lower than given in KDHK.

I expect from Tab. 1 of \cite{HVK} that this is the case because about
half of the intensity to the left of $Q_{\beta\beta}$ and about 20\%
of the peak at $2053 keV$ are predicted by the simulation.

\item ``There is no null hypothesis analysis demonstrating that the data
require a peak.''

The statement that there is a peak with probability $K$ implies  
that there may be none. Since the results are
probabilistic, it is not possible to demonstrate that
the data do require a peak.

Note that the model used to analyze the data, includes the possibility
that the intensity of the line is zero. The error intervals given in KDHK
have been chosen such as to exclude the value of zero. Therefore
every result in KDHK can especially be read as ``With probability
$K$ the intensity zero is outside the error interval''. In this sense
the null hypothesis is rejected with the probability onto which the 
error interval is based.

\item ``There is no statement of the net count rate of the peaks other
than the 2039-keV peak''. 

See item 2.

\item ``There is no presentation of the entire spectrum ...''

The entire spectrum was published \cite{HDM} shortly before KDHK.
See, however, item 2 for the normalization of Fig. 1 of \cite{HDM}.

\item ``There are three unidentified peaks in the region of analysis
that have greater significance than the 2039-keV peak ...''

Since they have higher significance than the significance claimed for the 
peak at $Q_{\beta\beta}\, ,$ even future improved analyses need not 
necessarily consider these peaks to be part of the background.
In principle there is no reason to consider all non-identified
structures as fluctuations of the background. Peaks that have a high 
significance may be considered a spectral line although unknown 
at present. 

Hence, a certain arbitrariness of the results remains even when the
size of the analysis window is beyond question.

\item ``There is no discussion of the relative peak strengths before and after
the single-site-event cut ...''

The intensity at $Q_{\beta\beta}$ after the single-site cut 
is compatible with the intensity before the cut, if the efficiency of
the single site cut is taken into account.

It is due to the choice of the analysis window that this question has not been
discussed for the other peaks.

\item ``No simulation has been performed to demonstrate that the
analysis correctly finds true peaks or that would find no peaks
if none existed ...''

This cannot be demonstrated. The randomness of the
data entails probabilistic conclusions. Especially there is no 
analysis that would find only the true peaks and would never confuse 
a fluctuation with a peak. For this reason the results of KDHK have 
been expressed in terms of the probability that the intensity of the 
peak lies in a given interval. 

KDHK had verified by Monte Carlo simulations of the analysis
(not to be confused with the simulation of the experimental setup
published in \cite{HVK}) that their statements are 
approximately right if their model is right: The true value of the inferred 
parameter lies with the relative frequency $K$ in the Bayesian error interval 
${\cal B}(K)\, .$ The Monte Carlo simulations of the analysis 
had been done to make sure
that no numerical errors are present. The simulations have not been mentioned
by KDHK because a mathematical theorem ensures that one comes at least
close to the above frequency interpretation of the
error interval. It is known from the work of Stein
\cite{Stein:65} and Villegas \cite{Villegas:77b,Chang:86} and others
that --- under certain symmetry conditions --- Bayesian error intervals 
have the desired frequency interpretation. In the present case, the symmetry
conditions are not exactly fulfilled because the parameter --- the
intensity --- is real and the observed events are natural numbers. 

\item ``There is no discussion of how sensitive the conclusions are to 
different mathematical models ...''

Unfortunately, any results --- not only the ones of KDHK --- are
sensitive to the model one  chooses
to describe peak plus background. The comparison between parts (a)
and (b) of Figs.4--6 in KDHK shows it --- since the size of the analysis
window is part of the model. The model dependence will not disappear
when the window is chosen wider. The result continues to depend on the
parameterization of background and contaminating peaks.

The model dependence is thus not the problem of KDHK only. 
It is a ubiquitous problem. Therefore the available methods of 
statistical inference may need improvement.

What one would like to have, is a procedure to infer the intensity of
the peak in question independently of the ansatz for the background. 
This ansatz should extend over an 
undisputable length and it should be sufficiently general so that
details like the constancy of the background do not matter. 
Such a procedure is not available at present.
The problem of the ``marginalisation paradox'' \cite{Dawid:73} gives a 
hint how to proceed: Independent parameters are those unaffected by it. 
Once the notion of independence is clarified, one can devise a
procedure to determine the intensity of the interesting line
(to a certain extent) independently of the assumptions about the
background. This is being worked out.

\end{enumerate}

In summary: The critics have made a valuable point with their concern
about the size of the analysis window; the simulation of the
experimental setup \cite{HVK} indicates that the significance of the
possible strucuture at $Q_{\beta\beta}$ is lower than claimed by KDHK.

\end{document}